\documentclass[pre,twocolumn,showpacs]{revtex4}

\RequirePackage{times}
\usepackage{graphicx, color}
\usepackage{latexsym}
\usepackage{dcolumn}
\usepackage{bm}

\newcommand{\D}{\displaystyle}

\begin{document}
\title{Experimental Evidence of the Role of Compound Counting Processes\\ in Random Walk Approaches to Fractional Dynamics}
\author{Justyna Trzmiel}\email{justyna.trzmiel@pwr.wroc.pl}\author{Karina Weron}\email{Karina.Weron@pwr.wroc.pl}
\affiliation{Institute of Physics,\\ Wroc{\l}aw University of Technology\\Wyb.Wyspia{\'n}skiego 27,
50--370 Wroc{\l}aw, Poland}%
\author{Aleksander Stanislavsky}\email{alexstan@ri.kharkov.ua}
\affiliation{Institute of Radio Astronomy, 4 Chervonopraporna St., 61002 Kharkov,
Ukraine}
\author{Agnieszka Jurlewicz}\email{Agnieszka.Jurlewicz@pwr.wroc.pl}
\affiliation{Hugo Steinhaus Center, Institute of Mathematics and Computer Science,\\ Wroc{\l}aw University of Technology\\Wyb.Wyspia{\'n}skiego 27,
50--370 Wroc{\l}aw, Poland}%

\begin{abstract}
We present dielectric spectroscopy data obtained for gallium-doped
Cd$_{0.99}$Mn$_{0.01}$Te:Ga mixed crystals which exhibit a very
special case of the two-power-law relaxation pattern with the
high-frequency power-law exponent equal to 1. We explain this
behavior, which cannot be fitted by none of the well-known
empirical relaxation functions, in a subordinated diffusive
framework.  We propose diffusion scenario based on a renormalized
clustering of random number of spatio-temporal steps in the
continuous time random walk. Such a construction substitutes the
renewal counting process, used in the classical continuous time
random walk methodology, by a compound counting one.  As a result,
we obtain a novel relaxation function governing the observed
non-standard pattern, and we show the importance of the compound
counting processes in studying fractional dynamics of complex
systems.
\end{abstract}


\maketitle

\section{Introduction}
No doubt, the linear response theory \cite{Kubo} plays a great
role in  physics, being a theoretical background and fundamental
accomplishments of statistical physics. Its conventional
formulation is based on two assumptions \cite{Allegrini}: 1) The
time evolution of the system variables is governed by Hamiltonian
operators; 2) The external perturbation arising from an initial
disturbance makes the system depart from canonical equilibrium
weakly, and the linear response function is expressed in terms of
the derivative of a stationary correlation function (of dipole
orientation in the theory of relaxation, for example). This gives
a rigorous and efficient approach to the temporal description of
some Hamiltonian systems \cite{Lee}. However, the foundations are
crashed on many-body kernel in dynamics of complex systems. The
main cause is that the macroscopic evolution of such systems
cannot be attributed to any particular object chosen from those
forming the systems. Any form of many-body interactions in the
complex systems should be introduced through cross-correlations
between their different objects, yielding hence insurmountable
mathematical problems. Therefore, it is very difficult, if
possible generally, to describe the time evolution of many-body
systems by Hamiltonian operators, and any stationary correlation
function is not available at all. Since, in general, the dynamical
processes in complex systems are strictly stochastic in nature,
they should be analyzed in a corresponding manner.

For description of relaxation and transport properties in such
systems as glasses, liquid crystals, polymers, etc., the
continuous-time random walk (CTRW) processes are one of the most
useful mathematical tools. Despite their long history, started
with the brilliant Montroll-Weiss idea \cite{Montroll}, the CTRWs
are still far from their full exploration. However, their
connection with anomalous diffusion has been already recognized
(see e.g., \cite{Metzler}). Recently, a progress in understanding
of this mathematical tool \cite{Meerschaert1,  Meerschaert2,
Piryatinska, Jurlewicz, Magdziarz} stimulated new developments in
diffusive scenarios of the non-exponential relaxation phenomena
\cite{WeronA, Weron1, Stanislavsky}.

In this paper we present experimental data which confirm that the
non-exponential relaxation behavior, in fact, is governed by
compound counting processes strictly connected with clustered
CTRWs \cite{Jurlewicz}. In Sec.~\ref{part2}, we study dielectric
spectroscopy data measured for gallium-doped
Cd$_{0.99}$Mn$_{0.01}$Te:Ga semiconducting mixed crystals
possessing deep, metastable defects. We observe that this material
exhibits such a relaxation pattern which cannot be fitted with any
of the well-known empirical relaxation functions. Hence, we apply
a novel relaxation law  being a modification of the result derived
recently in \cite{Stanislavsky}. In Sec.~\ref{part3} we propose an
anomalous diffusion scenario, based on  the notion of a compound
counting process, by means of which the observed relaxation
behavior may be explained. We end with conclusions in
Sec.~\ref{part4}.

\section{Experiment}\label{part2}

Dielectric spectroscopy studies carried out on various physical
systems revealed that a wide class of materials follows the
anomalous relaxation mechanism \cite{Jonscher1} represented by
low- and high-frequency fractional power-law dependences of the
imaginary part $\varepsilon''(\omega)$ of the complex dielectric
permittivity
$\varepsilon^{*}(\omega)=\varepsilon'(\omega)-i\varepsilon''(\omega)$:
\begin{equation}
\label{power_law}
\begin{array}{l l}
\varepsilon''(\omega) \sim (\omega / \omega_p)^m,
&
{\omega} \ll \omega_{p},\\
\varepsilon''(\omega) \sim (\omega / \omega_p)^{n-1},& \omega \gg \omega_{p},
\end{array}
\end{equation}
where $\omega_p$ denotes the loss peak frequency and
$0<m,n<1$. Depending  on a mutual relation between the power-law
exponents two different types of relaxation responses can be
distinguished. The relaxation response is called typical when the
power-law exponents satisfy relation $m\ge 1-n$. In order to
interpret this type of experimental data the well-know
Havriliak-Negami (HN) function \cite{Jonscher1,Jonscher2}

\begin{equation}
\label{HN} \varphi _{HN}^{*}\left( \omega  \right)= \D{1\over
\left[1+\left(i\omega/\omega_p\right)^{\alpha}\right]^{\gamma}},
\;\;\;\;\;0<\alpha,\gamma<1
\end{equation}
is used: ${{\varepsilon }^{*}}\left( \omega  \right)= \left(
{{\varepsilon }_{0}}-{{\varepsilon }_{\infty }}
\right){{\varphi_{HN} }^{*}}\left( \omega  \right)+{{\varepsilon
}_{\infty }}$, where  ${{\varepsilon }_{0}}$ is the static
permittivity and ${{\varepsilon }_{\infty}}$   represents the
asymptotic value of the dielectric permittivity at high
frequencies. For the HN function (\ref{HN}) the two-power-law
property (\ref{power_law}) is fulfilled with the power-law
exponents $m=\alpha$ and $1-n=\gamma\alpha$. For a long time
period the HN function with extended parameters' range $0<\alpha,
\alpha\gamma <1$ was also used to fit the less typical relaxation
data for which the power-law exponents yield the opposite
inequality $m<1-n$. Unfortunately, none of the known relaxation
models \cite{Weron1,Kalmykov} can justify the values $\gamma >1$
appearing in the extended range of the power-law exponents. Recent
progress in stochastic modeling of relaxation processes has
resulted in derivation of a new relaxation pattern \cite{Weron1,
Stanislavsky} underlying the less typical responses:

\begin{equation}
\label{JWS} \varphi^{*}\left( \omega  \right)= 1-\D{1\over
\left[1+\left(i\omega/\omega_p\right)^{-\alpha}\right]^{\gamma}},
\;\;\;\;\;0<\alpha,\gamma<1.
\end{equation}
This function exhibits the two-power-law property
(\ref{power_law})  with the power-law exponents $m=\alpha\gamma$
and $1-n=\alpha$. It not only properly describes the less typical
class of relaxation responses but also relates the experimentally
observed power-law properties with random characteristics of the
investigated system. Let us notice that both formulas (\ref{HN})
and (\ref{JWS}) can be used as fitting functions also with
parameters $\alpha=1$ and/or $\gamma=1$. In such cases they result
from slightly different diffusion scenarios and may exhibit one
fractional power-law only or even none.

The less typical relaxation pattern is observed in semiconducting
mixed crystals of $\rm Cd_{0.99}Mn_{0.01}Te$:Ga possessing deep
metastable defects -- the so called DX centers. It is widely
accepted that depending on the position of gallium (Ga) dopants in
the CdMnTe lattice shallow donor states or deep metastable traps
may be formed \cite{Park}. As shown in Ref. \cite{Trzmiel1} the
relaxation response of $\rm Cd_{0.99}Mn_{0.01}Te$:Ga is mainly
influenced by presence of deep, metastable traps within the band
gap of the investigated $\rm \mbox{\rm Au-Cd}_{0.99}Mn_{0.01}Te$
Schottky junction. For purpose of the present study we analyzed
the frequency-domain response of two samples of the same x = 0.01
manganese (Mn) content labeled as sample 1 and sample 2,
respectively. Both the samples posses the same net donor
concentration of approximately $10^{15}$ $\rm cm^{-3}$ estimated
from the capacitance-voltage measurements. Description of the
samples preparation can be found in details elsewhere
\cite{Trzmiel2}. Gold Schottky contacts were thermally evaporated
on the front side of the samples. Measurements were performed at
zero bias using Novocontrol impedance analyzer. Applied ac probe
signal amplitude was equal to 10 mV.

\begin{figure}[htbp]
\includegraphics[scale=0.68]{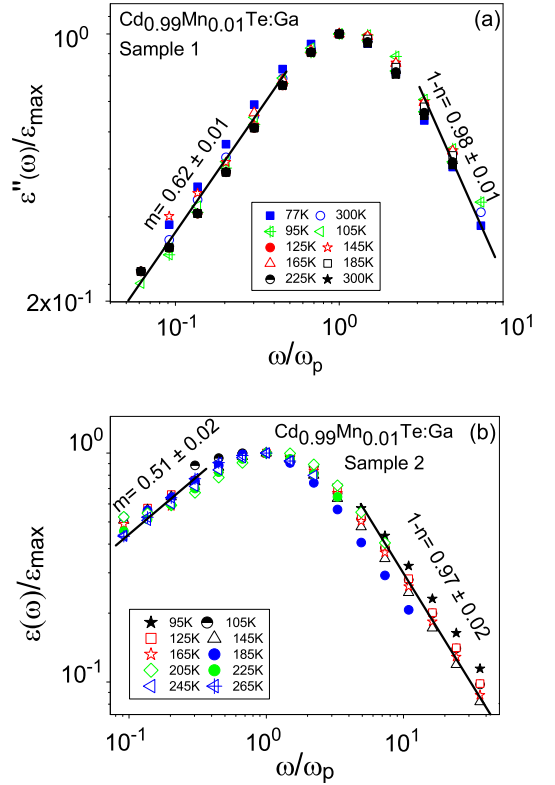}
\caption{Normalized imaginary part of dielectric permittivity
obtained for two samples of $\rm Cd_{0.99}Mn_{0.01}Te$:Ga. Both
the samples exhibit two-power-law relaxation behavior with
power-law exponents satisfying relaxation $m<1-n$.}
\label{figure1}
\end{figure}

In Figure~\ref{figure1} the normalized imaginary part of the
dielectric permittivity for two samples, investigated in a broad
temperature range, is presented. It is clear from the plot that
both the samples exhibit the less typical, two-power-law
relaxation pattern $m<1-n$, however, with different values of the
low-frequency power-law exponent $m$. The change in values of this
exponent, in the samples of the same manganese content and
gallium concentration, may be associated with locally different
surroundings of DX centers contributing to the effective
relaxation response of the investigated sample.


\begin{figure}[htbp]
\includegraphics[scale=0.8]{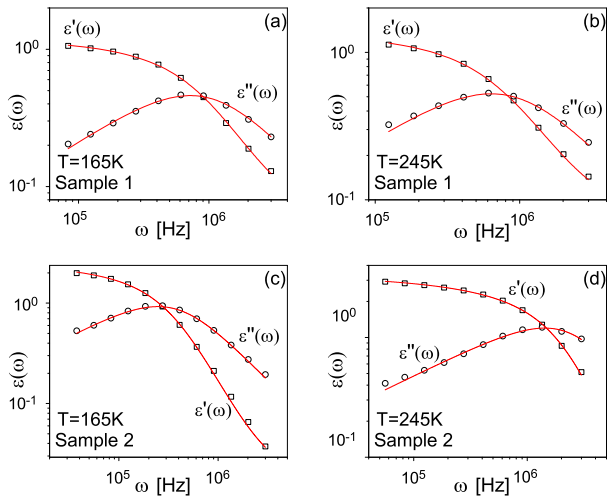}
\caption{Sample real and imaginary part of the permittivity data
as a function of frequency obtained for the two investigated
samples of $\rm Cd_{0.99}Mn_{0.01}Te$:Ga  at various temperatures.
Solid lines represent function (\ref{JWS}).} \label{figure2}
\end{figure}

\begin{figure}[htbp]
\includegraphics[scale=0.6]{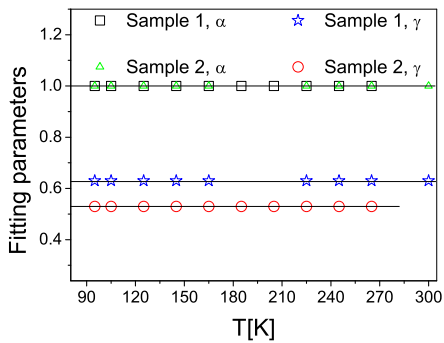}
\caption{The values of fitting  parameters obtained for two
samples of $\rm Cd_{0.99}Mn_{0.01}Te$:Ga obtained at various
temperatures. Sample 1 : $\alpha =1.00\pm 0.01$ and $\gamma
=0.63\pm 0.01$;  Sample 2 : $\alpha =1.00\pm 0.01$ and $\gamma
=0.53\pm 0.01$.} \label{figure3}
\end{figure}

In Figure~\ref{figure2} the experimental data fitted by means of
formula (\ref{JWS}) are presented. It can be observed that this
function perfectly covers the experimental data points.  Values of
the fitting parameters are collected in Figure~\ref{figure3}. Both
$\alpha $ and $\gamma $ values are temperature independent. It
should be pointed out that in case of both analyzed samples
$\alpha $ parameter remains the same, whereas value of $\gamma$ is
significantly different depending on the sample considered. Since
$\alpha$  is approximately equal to~1, we observe lack of the
high-frequency fractional power-law behavior. Such a relaxation
pattern, being a very special case of (\ref{power_law}), suggests
a non-standard diffusion scenario underlying the experimental
result.

\section{Diffusion scenario}\label{part3}
The CTRW process $R(t)$ determines the total distance reached by a
random walker until time $t$. It is characterized by a sequence of
independent and identically distributed (i.i.d.) spatio-temporal
random steps $(R_i,T_i), i\geq 1$. If we assume stochastic
independence  between jumps $R_i$ and waiting times $T_i$, we get
a decoupled random walk; otherwise we deal with a coupled CTRW.
The distance reached by the walker at time $t$ is given by the
following  sum
\begin{equation}
\label{diff_ctrw_general}
R(t)=\sum^{\nu(t)}_{i=1}R_i\,,
\end{equation}
where $\nu(t)=\max\{n:\,\sum_{i=1}^n T_i\leq t\}$, counts the
performed steps.

Theoretical studies of the relaxation phenomenon in the above
framework are based on the idea of an excitation undergoing
(anomalous, in general) diffusion in the system  under
consideration \cite{Metzler}. The relaxation function $\phi(t)$ is
then defined by the inverse Fourier or the Laplace transform of
the diffusion front $\bar R (t)\approx R(t/\tau_0)/f(\tau_0)$,
where the dimensionless rescaling parameter $\tau_0\approx 0$ and
$f(\tau_0)$ is appropriately chosen renormalization function. The
diffusion front $\bar R (t)$ approximates a position at time $t$
of the walker performing rescaled spatio-temporal steps
$(R_i/f(\tau_0),\tau_0 T_i)$. The characteristics of the
relaxation process are related to the properties of the diffusion
front resulting from assumptions imposed on the spatio-temporal
steps of the random walk. For example, the decoupled CTRW with
power-law waiting-time distributions (i.e., with the random
variables $T_i$ satisfying ${\rm Prob}(T_i\geq t)\sim(t/t_0)^{-a}$
as $t\to\infty$ with some $0<a<1$ and $t_0>0$) leads to the
Cole-Cole relaxation \cite{Kotulski2}. However, the
frequency-domain Cole-Cole relaxation with the corresponding
time-domain Mittag-Leffler pattern is only one of the cases
measured in various experiments with complex media, and derivation
of those more general patterns requires considering diffusion
scenarios based on a compound coupled CTRW representation. It
should be pointed out that the simple coupling of type $R_i\sim
T^p_i$ (with positive power exponent $p$) does not lead behind the
Cole-Cole relaxation \cite{Kotulski1}. In contrast, introducing a
dependence between the jumps and waiting times by a random
clustering procedure we can obtain another empirical relaxation
laws like the Cole-Davidson or Havriliak-Negami patterns
\cite{JurlewiczA,Weron1,Stanislavsky}. Below, we present the
diffusion scenario which leads directly to the results discussed
in the preceding section.

Let $M_j$ be a sequence of i.i.d. positive integer-valued random
variables independent of the pairs $(R_i,T_i)$. Next, assume that
the jumps and  waiting times are assembled into clusters of random
sizes $M_1,M_2,\dots$. This assumption allows one to transform the
sequence of spatio-temporal steps $(R_i,T_i)$ into a new sequence
$(\tilde{R_j},\tilde{T_j})$ of random sums
\begin{eqnarray}
\label{clustering}
(\tilde{R_1},\tilde{T_1})&=&\sum_{i=1}^{M_1}(R_i,T_i)\,,\nonumber\\
(\tilde{R_j},\tilde{T_j})&=&\sum_{i=M_1+\dots+M_{j-1}+1}^{M_1+\dots+M_j}
(R_i,T_i)\,,\quad j\geq 2.
\end{eqnarray}
Then the position $R^M(t)$ of the walker is determined by
$(\tilde{R_j},\tilde{T_j})$ and, in accordance with the general
formula (\ref{diff_ctrw_general}), it is given by
\begin{equation}
\label{cluster1}
R^M(t)=\sum^{\tilde\nu(t)}_{j=1}\tilde R_j\,,
\end{equation}
where $\tilde\nu(t)=\max\{n:\,\sum_{j=1}^n\tilde T_j\leq t\}$. The
dependence between the jumps $\tilde R_j$ and the waiting times
$\tilde T_j$ of the coupled CTRW process $R^M(t)$ is determined by
the distribution of the cluster sizes $M_j$ \cite{Jurlewicz}.

In the simple case when the waiting times are represented
by equal intervals in time, i.e., $T_i=\Delta t$ , we have
\begin{equation}
\label{diff_ctrw}
R(t)=\sum^{\lfloor t/\Delta t\rfloor}_{i=1}R_i\,,
\end{equation}
where $\lfloor \cdot \rfloor$ denotes the integer part, while the
clustering procedure (\ref{clustering}) yields   $\tilde
T_j=M_j\Delta t$ for $j\geq 1$, and the coupled process $R^M(t)$
in  (\ref{cluster1}) takes an equivalent form
\begin{equation}
\label{M-CTRW}
R^M(t)=\sum^{U^M({\tilde\nu(t)})}_{i=1}R_i\,.
\end{equation}
Here $U^M(\tilde\nu(t))$ is a compound counting process obtained
from $U^M(n)=\sum_{j=1}^n \tilde T_j/\Delta t =\sum_{j=1}^n M_j$,
and $\tilde\nu(t)=\max\{n:\,\sum_{j=1}^n \tilde T_j\leq
t\}=\max\{n:\,U^M(n)\leq t/\Delta t\}$. Observe that formula
(\ref{M-CTRW}) is an analog of  (\ref{diff_ctrw}) with the
compound counting process $U^M(\tilde\nu(t))$ substituting the
deterministic number $\lfloor t/\Delta t\rfloor$ of performed
jumps $R_i$. The counting process $U^M(\tilde\nu(t))$ is always
less than $\lfloor t/\Delta t\rfloor$, and it is hence a special
case of the undershooting compound counting process \cite{Weron1}.
It is also a clear signature of  the spatio-temporal coupling
provided by the clustering procedure (\ref{clustering}).

The idea of compound counting processes in CTRW approach is not
new in physics.  The resulting CTRW processes were examined in the
context of the rareness hypothesis in the fractal-time random walk
models (see, e.g. \cite{Weissmann, Klafter}). In general, the
compound counting process cumulates random number of random
events. Physical situations where the relevance of this scheme
holds are numerous. For instance, take into account the energy
release of individual earthquakes in geophysics, the random
magnitude of claims' sequence in insurance risk theory or random
water inputs flowing into a dam in hydrology where summing the
individual contributions yields the total amount of the studied
physical magnitude over certain time intervals.

\begin{figure}
\includegraphics[scale=0.5]{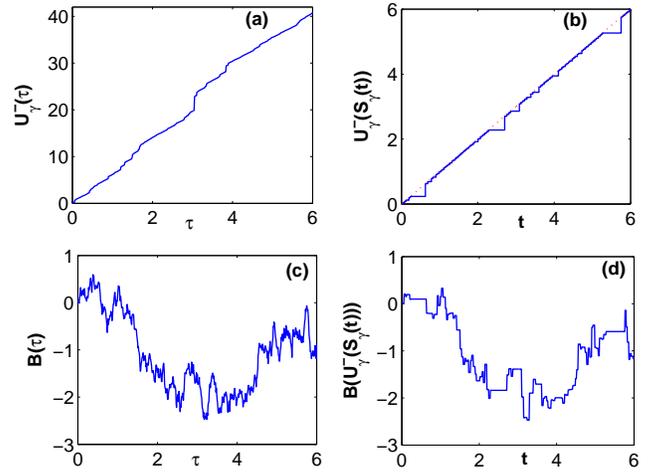}
\caption{(Color online) a) Left-continuous process
$U^{-}_\gamma(\tau)$; b) Compound counting process
$Z(t)=U^{-}_\gamma[S_\gamma(t)]$ as a new arrow of time; c) 1D
trajectory of standard Brownian motion (parent process); d) 1D
trajectory of Brownian motion subordinated by $Z_U(t)$. Here
everywhere the index $\gamma$ equals to 0.9.} \label{figure4}
\end{figure}

\begin{figure}
\includegraphics[scale=0.5]{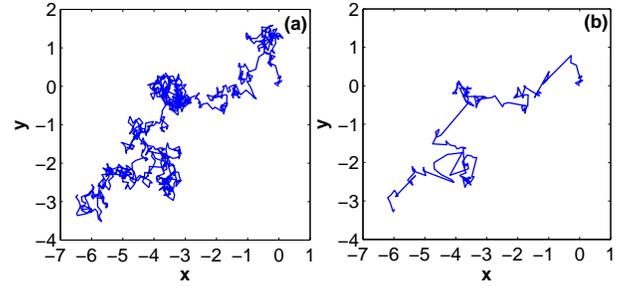}
\caption{(Color online) 2D trajectories of standard Brownian
motion (a) and Brownian motion under compound subordination with
$\gamma=0.7$ (b).} \label{figure5}
\end{figure}

The diffusion front $\bar R^M (t)$ related to (\ref{M-CTRW}) takes
the subordinated form
 \begin{equation}
\label{M-CTRW-diff-front}
\bar R^M(t)\stackrel{d}{=} X(Z(t))\,,
\end{equation}
where the parent process $X(\tau)$ is just the diffusion front
corresponding to the simple random walk $R(t)$ given by
(\ref{diff_ctrw}), while the undershooting subordinator $Z(t)$
corresponds to the limit of the rescaled counting process
$U^M(\tilde\nu(t/\tau_0))/g(\tau_0)$ as $\tau_0\to 0$ (with
renormalizing function $g(\tau_0)$ chosen appropriately). Both the
parent process and the undershooting subordinator are well defined
if the distributions of the spatial steps $R_i$ and cluster sizes
$M_j$ satisfy some conditions, referring to their asymptotic
behaviors. In particular, if  $\langle M_j\rangle<\infty$ (or
equivalently $\langle \tilde T_j\rangle<\infty$) we have
$Z(t)=t/\Delta t$. On the other hand, taking  into account
clustering with a heavy-tailed cluster-size distribution
\begin{displaymath}
{\rm Prob}(M_j\geq m)\mathop{\sim}_{m\to\infty}(m/c)^{-\gamma}
\end{displaymath}
with the tail exponent $0<\gamma<1$ and some scaling constant
$c>0$, we obtain the compound form
$Z(t)=U^-_\gamma[S_\gamma(t/\Delta t)]$, where
$U^-_\gamma(\tau)=\lim_{x\to\tau_-}U_\gamma(x)$ is the left limit
of the $\gamma$-stable subordinator $U_\gamma(\tau)$, and
$S_\gamma(t)=\inf\{t\geq0\,:\,U(\tau)>t\}$ is its inverse process.
For a fixed $t>0$, the compound undershooting subordinator
$U^-_\gamma[S_\gamma(t/\Delta t)]$ has the generalized arcsine
distribution rescaled by $t/\Delta t$. The corresponding
probability density function reads
\begin{equation}
p_\gamma(t,\tau)=\frac{\sin\pi\gamma}{\pi}\,\tau^{\gamma-1}
(t/\Delta t-\tau)^{-\gamma}\,,\quad0<\tau<t/\Delta t\,.
\end{equation}
It is easy to check that all moments of the random variable
$U^-_\gamma[S_\gamma(t/\Delta t)]$ are finite.

The numerical approximation of the process
$U^-_\gamma[S_\gamma(t/\Delta t)]$ is shown in Fig.~\ref{figure4}.
For simplicity, we take $\Delta t=1$. To simulate the process
$U^-_\gamma[S_\gamma(t)]$, one only needs to generate the values
$U_\gamma(n\Delta \tau)$ (see Fig.~\ref{figure4}(a)), where
$n=1,2,\dots$, and $\Delta \tau$ is the step length. This can be
carried out by the standard method of summing up the independent
and stationary increments of the L\'evy process \cite{JanWer94}.
Then the approximation of the process $U^-_\gamma[S_\gamma(n\Delta
t)]$ is a simple continuous-time random walk in which each waiting
time is exactly equal to the jump. In the result,
Fig.~\ref{figure4}(b) demonstrates the process
$U^-_\gamma[S_\gamma(t)]$ non-decreasing in time.

As far as the parent process is concerned, for simplicity, but
without loss of generality, we can consider  the standard Brownian
motion $B(t)$ that can easily be obtained if we take into account
the spatial jumps satisfying conditions $\langle R_i\rangle=0$ and
$\langle R_i^2\rangle<\infty$ (see Fig.~\ref{figure4}(c)). Then
the subordinated process $B(U^-_\gamma[S_\gamma(t)])$ represents
the corresponding diffusion front. Its sample paths are shown in
Fig.~\ref{figure4}(d). The main feature of the subordinated
process consists in random jumping along random trajectories of
the parent process. For better visualization of this feature we
can consider a walk on a plane with 2D vector jumps satisfying
analogous conditions that leads to 2D Brownian motion as a parent
process, see Fig.~\ref{figure5}. The difference between
Figs.~\ref{figure5}(a) and (b) is that the walker, moving along a
Brownian trajectory in presence of the compound subordination,
stops from time-to-time and overjumps through intermediate
positions in the Brownian trajectory. In other words, the
trajectory of Fig.~\ref{figure5}(b) is something like a partition
of the trajectory in Fig.~\ref{figure5}(a) on random intervals.
The appearance of the overjumps, as applied to the relaxation of
semiconductors with metastable defects described in Sec.
\ref{part2}, is connected with long-range interactions among DX
centers.

The probability density function of
$B(U^-_\gamma[S_\gamma(t/\Delta t)])$ reads
\begin{equation}
\label{diff_f_density}
p(x,t)=\int_0^\infty
p^B(x,\tau)\,p_\gamma(t,\tau)\,d\tau\,,
\end{equation}
where $p^B(x,\tau)={1\over\sqrt{2\pi\tau} } e^{-x^2/2\tau}$.
Taking the inverse Fourier transform with respect to $x$ for
$p(x,t)$ in Eq.(\ref{diff_f_density}), we derive the relaxation
function $\phi(t)=\langle e^{-ik\bar R^M(t)}\rangle$ that
characterizes the temporal decay of a given macroscopic mode $k$.
We get
\begin{equation}
\label{relax_funct}
\phi(t)=\frac{\sin\pi\gamma}{\pi}\int_0^{t/\Delta t}
e^{-(k^2/2)\tau}\,\tau^{\gamma-1}(t/\Delta t-\tau)^{-\gamma}\,d\tau\,,
\end{equation}
where $k^2/2$ may be denoted as a characteristic frequency
$\omega_p$ of the relaxing system. The integral representation of
the relaxation function may be substituted by a series expansion
observing that the relaxation function (\ref{relax_funct}) can be
expressed as $\phi(t)= E_{1,1}^\gamma(-\omega_p\,t)$,  where
\begin{displaymath}
E_{\alpha,\beta}^\gamma(x)=\sum_{k=0}^\infty
\frac{(\gamma,k)\,x^k}{\Gamma(k\alpha+\beta)k!}\,,\quad
\alpha,\beta>0\,,
\end{displaymath}
is the generalized Mittag-Leffler function \cite{MatSaxHaub09}.
Here $(\gamma,k)=\gamma(\gamma+1)(\gamma+2)\dots(\gamma+k-1)$  is
the Appell's symbol with $(\gamma,0)=1$, $\gamma\neq 0$.

Based on the Laplace image of $\phi(t)$ with respect to $t$ we
find  the kinetic equation of relaxation in the pseudodifferential
form
\begin{displaymath}
\left(\frac{d}{dt}+\omega_p\right)^\gamma\phi(t)=
\frac{t^{-\gamma}}{\Gamma(1-\gamma)}\,,
\end{displaymath}
with the initial condition $\phi(0)=1$. For the experimental
studies of the dielectric spectroscopy data (dielectric
permittivity or the corresponding susceptibility) the
frequency-domain representation of the latter function
\begin{equation}
\varphi^*(\omega)\ = \int^\infty_0e^{-i\omega
t}\,\left(-\frac{d\phi(t)}{dt}\right)\,dt
\end{equation}
is of interest. Taking the Fourier transform, we get

\begin{equation}
\label{JSW2}
\varphi^*(\omega)\ = 1-\Bigg(\frac{i\omega/\omega_p}
{1+i\omega/\omega_p}\Bigg)^\gamma\,,\quad 0<\gamma< 1\,.
\end{equation}
Let us note that formula (\ref{JSW2}) coincides with (\ref{JWS})
for $\alpha =1$. Moreover, for $\gamma=1$ it takes the form of the
Debye response, corresponding to the spatio-temporal clustering
with $\langle M_j\rangle<\infty$, where the undershooting
subordinator becomes deterministic: $Z(t)=t/\Delta t$.

Considering the dielectric susceptibility ${\chi^*(\omega)=
\chi'(\omega)- i{\chi''(\omega)}}=\chi(0)\varphi^{*}(\omega)$ with
$\varphi^*(\omega)$ as in (\ref{JSW2}) one can show that the real
and imaginary parts of the susceptibility fulfill the following
frequency-independent relations:
\begin{eqnarray}
&\lim\limits_{\omega\to 0}&\frac{\chi''(\omega)}{\chi'(0)-\chi'
(\omega)}=\tan\Big(\frac{\pi\gamma}{2}\Big)\,,\nonumber\\
&\lim\limits_{\omega\to
\infty}&\frac{\chi''(\omega)}{\chi'(\omega)}=\infty.
\end{eqnarray}
This means that for small $\omega$ the gallium-doped $\rm
Cd_{0.99}Mn_{0.01}Te$  mixed crystals obey the energy criterion
\cite{Jonscher2}, while for large $\omega$ the high-frequency
energy lost per cycle does not have a constant relationship to the
extra energy that can be stored by a static field. The similar
feature takes place in the Cole-Davidson law where, however, the
energy criterion is obeyed for large $\omega$ only. As it is well
known \cite{Jonscher2}, the presence of the universal law in the
relaxation in gallium-doped Cd$_{0.99}$Mn$_{0.01}$Te mixed
crystals is caused strictly by many-component interactions. This
effect is just accounted for using the compound counting random
processes.

\section{Conclusions}\label{part4}
In the diffusive framework we have brought to light the fractional
dynamics of gallium-doped Cd$_{0.99}$Mn$_{0.01}$Te:Ga mixed
crystals, for which the particular case of less typical relaxation
behavior with the high-frequency power-law exponent $1-n=1$ has
been detected. We have proposed the subordination scenario of
anomalous diffusion underlying the observed pattern, based on
random, heavy-tailed spatio-temporal clustering procedure applied
to the random walk hidden behind the standard Brownian motion.
Such a procedure has led to a substitution of the renewal counting
process, used in the classical CTRW model, by the undershooting
compound counting one. As a consequence, the explicit stochastic
structure of the diffusion front as a subordinated Brownian motion
has been obtained. The resulting frequency-domain relaxation
function has been shown to fit the studied dielectric spectroscopy
data.

\section{Acknowledgements}

Work of J.T. and A.J. was partially supported by project PB NN 507503539.

A.S. is much obliged to the Institute of Physics and the Hugo
Steinhaus Center for pleasant hospitality during his visit in
Wroc{\l}aw University of Technology.

J.T. is grateful to dr hab. Ewa Popko for making the samples of
Cd$_{0.99}$Mn$_{0.01}$Te:Ga accessible for measurements.

A.S. thanks dr Marcin Magdziarz for his useful remarks.



\begin{thebibliography}{2010}

\bibitem{Kubo} R. Kubo, M. Toda, and N. Hashitsume,  {\it Statistical Physics II}, (Springer, Berlin, 1985).

\bibitem{Allegrini} P. Allegrini, M. Bologna, L. Fronzoni, P. Grigolini, and L. Silvestri, Phys.~Rev.~Lett. {\bf 103}, 030602 (2009).

\bibitem{Lee} U. Balucani, M. H. Lee, and V. Tognetti, Phys.~Rep. {\bf 373}, 409 (2003).

\bibitem{Montroll} E.W. Montroll and G.H. Weiss, J.~Math.~Phys. {\bf 6}, 167 (1965).

\bibitem{Metzler} R. Metzler and J. Klafter, Phys.~Rep.~{\bf 339}, 1~(2000).

\bibitem{Meerschaert1} M.M.~Meerschaert, D.A.~Benson, H.-P.~Scheffler, and P.~Becker-Kern,
Phys.~Rev.~E~{\bf 66}, 060102(R) (2002).

\bibitem{Meerschaert2} M.M.~Meerschaert and H.-P.~Scheffler,
J.~Appl.~Probab.~{\bf 41}, 623~(2004).

\bibitem{Piryatinska} A.~Piryatinska, A.I.~Saichev, and W.A.~Woyczynski,
Physica~A~{\bf 349}, 375~(2005).

\bibitem{Jurlewicz} A.~Jurlewicz,
Diss.~Math.~{\bf 431}, 1~(2005).

\bibitem{Magdziarz} M.~Magdziarz and K.~Weron,
Physica~A~{\bf 367}, 1~(2006).

\bibitem{WeronA} A.~Weron and M.~Magdziarz,
EPL~{\bf 86}, 60010~(2009).

\bibitem{Weron1} K.~Weron, A.~Jurlewicz, M.~Magdziarz, A.~Weron, and J.~Trzmiel,
Phys.~Rev.~E~{\bf 81}, 041123~(2010).

\bibitem{Stanislavsky} A.A.~Stanislavsky, K.~Weron, and J.~Trzmiel,
EPL~{\bf 91}, 40003~(2010).

\bibitem{Jonscher1} A.K.~Jonscher, {\it Dielectric Relaxation in Solids},
(Chelsea Dielectrics Press, London, 1983).

\bibitem{Jonscher2} A.K.~Jonscher, {\it Universal Relaxation Law},
(Chelsea Dielectrics Press, London, 1996).

\bibitem{Kalmykov} Y.P.~Kalmykov, W.T.~Coffey, D.S.F.~Crothers, and S.V.~Titov,
Phys.~Rev.~E~{\bf 70}, 041103~(2004).

\bibitem{Park} C.H.~Park and D.J.~Chadi , Phys.~Rev.~B, {\bf 52}, 11884 (1995).

\bibitem{Trzmiel1} J.~Trzmiel, E.~Placzek-Popko, E.~Zielony, and Z.~Gumienny, Acta~Phys.~Pol. A {\bf 116}, 956~(2009).

\bibitem{Trzmiel2} J.~Trzmiel, K.~Weron, and E.~Placzek-Popko,
J.~Appl.~Phys.~{\bf 103}, 114902~(2008).

\bibitem{Kotulski2} K.Weron and M. Kotulski, Physica A~{\bf 232}, 180~(1996).

\bibitem{Kotulski1} M. Kotulski, in \textit{Chaos — The Interplay Between Stochastic and Deterministic
Behaviour}, edited by P. Garbaczewski, M. Wolf, and A. Weron, Lect. Notes
Phys. Vol. 457, Springer, Berlin 1995, p. 471.

\bibitem{JurlewiczA} A.~Jurlewicz and K.~Weron,
Acta~Phys.~Pol.~{\bf 39}, 1055~(2008).

\bibitem{Weissmann} H.~Weissmann, G.H.~Weiss, and S.~Havlin,
J.~Stat.~Phys.~{\bf 57}, 301~(1989).

\bibitem{Klafter} J.~Klafter and M.F.~Schlesinger,
J~Phys.~Chem.~{\bf 98}, 7366~(1994).



\bibitem{JanWer94}
A. Janicki and A. Weron, {\it Simulation and Chaotic Behaviour of
$\alpha$-Stable Stochastic Processes} (Marcel Dekker, New York,
1994).

\bibitem{MatSaxHaub09}
A.M. Mathai, R.K. Saxena, and H.J. Haubold, {\it The H-Function.
Theory and Applications} (Springer, Amsterdam, 2009).

\end{thebibliography}
\end{document}